\documentclass[a4paper,11pt]{article}
\pdfoutput=1 
             \usepackage{amssymb}
             \usepackage[intlimits]{amsmath}

\usepackage{jcappub} 

\usepackage[T1]{fontenc} 

\title{\boldmath Saving CNB assisted EDE model at the expense of quantum corrections?}

\author{Michael~Maziashvili}

\affiliation{School of Natural Sciences and Medicine, Ilia State University, 3/5 Cholokashvili Ave., Tbilisi 0162, Georgia}

\emailAdd{maziashvili@iliauni.edu.ge}

\abstract{As it is suggested in \cite{Sakstein:2019fmf, CarrilloGonzalez:2020oac}, one can dynamically introduce the coincidence time-scale for EDE in the framework of a particular mass-varying-neutrino-model as a time at which neutrinos constituting the cosmic neutrino background enter the non-relativistic regime. The model does not predict, however, the right amount of EDE density because of smallness of neutrino masses. One may hope to adjust the parameters in such a way as to ensure that the two-loop contributions are kept small while at the same time the effective mass for scalar field that enters the expression of zero-point-energy (for the field trapped in the minimum of effective potential) is sufficient for explaining the needed amount of EDE. Unfortunately, the answer is not in the affirmative.  }

\begin{document}
\maketitle
\flushbottom

\section{Introduction.}

Interestingly enough, the Hubble tension problem related to the discrepancy between the Hubble parameter values inferred from the measurements of cosmic microwave background (CMB) \cite{Planck:2018vyg} and the measurements utilizing type Ia supernovae calibrated with parallax and then Cepheid variable stars \cite{Riess:2019cxk} can be successfully resolved by admitting the presence of early dark energy (EDE) \cite{Poulin:2018dzj, Agrawal:2019lmo, Smith:2019ihp}. The role of EDE is to slightly increase the expansion rate of the universe prior to the recombination, while leaving the late universe unaltered. In particular, the contribution of EDE to the total energy density of the universe reaches about $10\%$ around the matter-radiation equality. Close to the recombination epoch, EDE should start to decay at least as fast as the radiation \cite{Poulin:2018cxd}. Modeling EDE with a scalar filed $\phi$, one may consider its coupling to matter fields in order to obtain a "dynamical explanation" of the coincidence $\rho_\phi/\rho_{tot}\simeq 0.1$ around the matter-radiation equality. One more logical issue is the explanation of quick "disappearance" of EDE after this coincidence (close to the recombination epoch). For addressing these issues one may find it useful to use mass-varying-neutrino-model \cite{Fardon:2003eh, Peccei:2004sz, Wetterich:2007kr, Amendola:2007yx, Brookfield:2005td, Brookfield:2005bz, MohseniSadjadi:2017pne} which is described by the equations of motion

\begin{eqnarray}
&& \dot{\rho}_\nu + 3H(\rho_\nu +p_\nu) =  \frac{\mathrm{d}\ln m_\nu}{\mathrm{d}\phi}(\rho_\nu - 3p_\nu)\dot{\phi} ~, \label{continuity} \\ && 
\ddot{\phi} +3H\dot{\phi} +U'(\phi) = -\sum_j\frac{\mathrm{d}\ln m_\nu}{\mathrm{d}\phi} (\rho_\nu - 3p_\nu) ~, \label{eqofmot} \\ && H^2 = \frac{8\pi}{3M_P^2}\left(\rho_\phi +\rho_\nu+\rho_r +\rho_m \right) \label{Friedmann}~.
\end{eqnarray} (In what follows we shall use natural units: $c=\hbar=1$.) Looking at these equations, one may argue that because of $\rho_\nu - 3p_\nu$ factor in Eqs.(\ref{continuity}, \ref{eqofmot}) the activation (or injection) of EDE should occur when the temperature of cosmic neutrino background (CNB) becomes of the order of $m_\nu$, that is when CNB enters the non-relativistic regime. In general, though, this idea does not work and one needs to make some particular arrangement of the model \cite{Sakstein:2019fmf, CarrilloGonzalez:2020oac}. One may assume that in the relativistic regime the factor $\rho_\nu - 3p_\nu$ becomes negligibly small that makes scalar field and neutrinos practically uncoupled. To address the issue of quick disappearance of EDE shortly after the coincidence in the framework of this kind of models, one may use a particular class of models, for instance \cite{Maziashvili:2021mbm} 

\begin{eqnarray}\label{example1}
U(\phi) = V\left(1-\mathrm{e}^{-\alpha \phi^2/M^2_P}\right) ~, ~ m_\nu(\phi) = \mu_\nu\mathrm{e}^{-\beta \phi^2/M^2_P} ~, ~~~
\end{eqnarray} in which EDE appears as a result of spontaneous symmetry breaking that restores somewhat later \cite{MohseniSadjadi:2017pne, Kepuladze:2021tsb}. The symmetry braking in this model is not concurrent with the epoch of coincidence, it takes place at earlier stages, but the time-scale of EDE-disappearance can be dynamically introduced as time of symmetry restoration.

The model suggested in \cite{Sakstein:2019fmf, CarrilloGonzalez:2020oac} 

\begin{eqnarray}\label{example2}
U(\phi) = \frac{\lambda \phi^4}{4} ~, ~~~~~~ m_\nu(\phi) = \mu_\nu\mathrm{e}^{-\beta \phi/M_P} ~, ~~~
\end{eqnarray} can also be viewed as leading to dynamical symmetry breaking despite the fact that $\mathcal{Z}_2$ symmetry of the potential is explicitly broken by the coupling term. For instance one may assume that $\phi$ is initially "frozen" at $\phi=0$ because of Hubble friction until it gets shifted to the minimum of the effective potential when the force due to coupling with neutrinos gets comparable to this friction force. This broken symmetry lasts then "forever". One should take a little care this sort of initial condition to look natural during the inflationary epoch. For this purpose one may assume a standard (cold) inflationary scenario which leads to the dilution of matter that gets reproduced at later times when the field starts to oscillate around the origin after the exit of slow rolling regime. That is, having admitted the possibility of cold inflation, the coupling with neutrinos can be neglected during inflation and one may assume that the field has had enough time to settle in the minimum of $U(\phi)$.

Let us note that for addressing the issue of EDE disappearance shortly after the coincidence in the framework of mass-varying-neutrino-model, there is one more suggestion \cite{Gogoi:2020qif} based on the clumping of CNB when they enter non-relativistic regime, which is due to fifth force between the neutrinos \cite{Afshordi:2005ym, Kepuladze:2021tsb}.

\section{Effective potential at the tree-level.}

The temperature of CNB at which the coincidence of EDE takes place is of the order of $1$\,eV - much smaller than the decoupling temperature of neutrinos: $\simeq 2$\,MeV. For this reason one can use the phase space distribution function of free streaming neutrinos, see Appendix \ref{appendixA}. In favor of the model \cite{Sakstein:2019fmf, CarrilloGonzalez:2020oac}, one can assume a quasi-degenerate spectrum of neutrino masses (see sections 14 and 26 in \cite{ParticleDataGroup:2020ssz})

\begin{eqnarray}
m^0_1 \simeq m^0_2 \simeq m^0_3 \simeq 0.05 \, \text{eV} ~. \nonumber 
\end{eqnarray} The superscript $0$ stands for the value at the time of coincidence. Under this assumption the factor $\mathsf{g}$ in Appendix \ref{appendixA} is equal to $6$. Then, as it is explained in Appendix \ref{appendixA}, when neutrinos are relativistic - the equation of motion for the scalar field takes the form

\begin{eqnarray}
\ddot{\phi} +3H\dot{\phi} = -\partial_\phi \left(U(\phi) + \frac{m_\nu^2(\phi)T_\nu^2}{8}\right) ~. 
\end{eqnarray} The effective potential

\begin{eqnarray}\label{effpotential}
\mathfrak{U}(\phi, T_\nu) \equiv U(\phi) + \frac{m_\nu^2(\phi)T_\nu^2}{8} ~, 
\end{eqnarray} for the model \eqref{example2} has a minimum at $\phi_+$ which obeys the equation

\begin{eqnarray}\label{minimum2}
\lambda\phi_+^3 = \frac{\beta\mu_\nu^2 T_\nu^2}{4M_P}\mathrm{e}^{-2\beta\phi_+/M_P} ~. 
\end{eqnarray} The temperature of the CNB can be parameterized as

\begin{eqnarray}\label{temperature}
T_\nu = 1.7\times 10^{-4} (1+z) \, \text{eV}  ~, 
\end{eqnarray} implying that at the coincidence $T_\nu \simeq 1$\,eV. Multiplying Eq.\eqref{minimum2} by $\phi_+/4$ and taking into account that we need

\begin{eqnarray}
\rho_\phi(\phi_+) = \frac{\lambda \phi_+^4}{4}  = 0.1\rho^0_{tot} ~, \nonumber 
\end{eqnarray} one finds

\begin{eqnarray}\label{phiplus}
0.1\rho^0_{tot} = \frac{\beta\mu_\nu^2 T_\nu^2\phi_+}{16M_P}\mathrm{e}^{-2\beta\phi_+/M_P} = \frac{T_\nu^2  (m_\nu^0)^2}{16} \, \frac{\beta \phi_+}{M_P} ~.~~~
\end{eqnarray} For evaluating $\rho^0_{tot}$ let us note that in the radiation dominated universe, which is the case under consideration, the radiation energy density below the $\mathrm{e}^+ \mathrm{e}^-$ annihilation temperature ($0.2$\,MeV) (until the neutrinos become non-relativistic) can be written as\footnote{In this temperature range, the radiation
	content of the universe consists of $3$ neutrino species and the photon.}

\begin{eqnarray}\label{radiation}
\rho_r = \left(1 + 3\frac{7}{8}\left(\frac{4}{11}\right)^{4/3}\right)\rho_\gamma = 1.1 T^4_\gamma = 4.3 T^4_\nu ~, \nonumber 
\end{eqnarray} At matter-radiation equality, that is at $T_\nu \simeq 1$\,eV, one obtains $\rho^0_{tot} \simeq 2\rho_r \simeq 8.6 $\,eV$^4$ and the Eq.\eqref{phiplus} reduces to

\begin{eqnarray}
\frac{\beta \phi_+}{M_P} = \frac{16\times 0.1\times 8.6}{0.05^2} = 5504 ~. 
\end{eqnarray} This result is not encouraging of course as we need very large $\mu_\nu$ to obtain realistic neutrino masses after the shift of the field to $\phi_+$

\begin{eqnarray}\label{huge}
\mu_\nu = m^0_\nu \mathrm{e}^{5504} ~. \nonumber 
\end{eqnarray} Before the field displacement, the neutrino masses are given by $\mu_\nu$ and this value is of course out of any cosmological bounds on neutrino masses.

It might be worth mentioning that we are facing analogous problem for non-relativistic CNB

\begin{eqnarray}
\mathfrak{U}(\phi, T_\nu) = \frac{\lambda\phi^4}{4} +  \frac{9\zeta(3) \mu_\nu T_\nu^3}{2\pi^2 }\, \mathrm{e}^{-\beta \phi/M_P}  ~. \nonumber 
\end{eqnarray} The minimum is achieved at 

\begin{eqnarray}
\lambda\phi_+^3   = \frac{9\zeta(3) \mu_\nu T_\nu^3}{2\pi^2 } \,\frac{\beta}{M_P} \, \mathrm{e}^{-\beta \phi_+/M_P} ~, \nonumber 
\end{eqnarray} which after multiplying by $\phi_+/4$ and setting $T_\nu = 1$\,eV takes the form

\begin{eqnarray}
\rho^0_\phi \,=\, 0.1\times 8.6\text{eV}^4 =  \frac{9\zeta(3) m_\nu^0 \text{eV}^3}{8\pi^2 } \,\frac{\beta\phi_+}{M_P} ~\Rightarrow \nonumber \\ \frac{\beta\phi_+}{M_P}  = \frac{8\pi^2 \times 0.1\times 8.6}{9\zeta(3)\times 0.05} \approx 126 ~.\nonumber 
\end{eqnarray} This result is as worse as Eq.\eqref{huge} as it would imply the neutrino masses prior to the coincidence epoch of the order of

\begin{eqnarray}
\mu_\nu =  \mathrm{e}^{126} 0.05\,\text{eV} ~. \nonumber 
\end{eqnarray} Thus, demanding a needed amount of EDE in model \eqref{example2} at the tree-level, we are facing not merely a parameter fine-tuning problem \cite{deSouza:2023sqp} but much more serious problem of going far beyond the neutrino mass bounds. Summarizing, a typical problem of mass varying neutrino models is that they can hardly provide the needed amount of EDE at the tree-level because of smallness of neutrino masses. That is because one has $\mu_\nu \simeq m_\nu^0 \Rightarrow \beta\phi_+/M_P \lesssim 1$ that allows one to readily evaluate the amount of EDE predicted by the model, see Eq.\eqref{phiplus}. It is worth noting that similar observation is also made on the 4th page of \cite{Sakstein:2019fmf}: "Interestingly, our proposal is on the verge of being excluded, and may even be so already".

It is easy to check that the above problem persists for renormalizable model

\begin{eqnarray}\label{example3}
U(\phi) = \frac{\lambda \phi^4}{4} ~, ~~~~ m_\nu(\phi) = \mu_\nu \left(1 - \frac{\beta \phi }{M_P}\right) ~,
\end{eqnarray} as well. The effective potential \eqref{effpotential} achieves its minimum at $\phi_+$ that obeys the equation

\begin{eqnarray}\label{minimum}
\lambda\phi_+^3 = \frac{T_\nu^2\mu_\nu m_\nu(\phi_+)}{4} \, \frac{\beta}{M_P} ~. 
\end{eqnarray} This equation tells us that

\begin{eqnarray}
\rho_\phi(\phi_+) = \frac{T_\nu^2\mu_\nu m_\nu(\phi_+)}{16} \, \frac{\beta\phi_+ }{M_P} =   \frac{T_\nu^2\big[\mu_\nu - m_\nu(\phi_+)\big] m_\nu(\phi_+)}{16}  ~, 
\end{eqnarray} that allows to make a crude order of magnitude estimate of EDE provided by the model as $(T^0_\nu m_\nu^0)^2 \simeq 0.0025$\,eV$^4$. It is obviously smaller than $0.86$\,\,eV$^4$ that we wanted to obtain. During the $\phi = 0$ stage the neutrino mass is given by $\mu_\nu$. Then it gets somewhat smaller when the field gets shifted to $\phi_+$ and after a while gets again dragged to $\mu_\nu$ since the CNB temperature decays and the minimum of effective potential \eqref{effpotential} moves towards $\phi=0$. Again we see that because of smallness of neutrino masses - the mass varying neutrino model for EDE does not work at the tree-level. This is also the case for a particular model given by Eq.\eqref{example1} \cite{Maziashvili:2021mbm}. In \cite{Maziashvili:2021mbm} it was  attempted to save the model by considering the one-loop  quantum corrections to $\rho_\phi$. However, the renormalization procedure to the model \eqref{example1} can be applied only formally, while one needs to be really careful in evaluating the validity conditions for restricting oneself by the one-loop approximation to the
quantum corrections.

	\section{Quantum corrected Effective potential.}
	
	It may happen that the radiative corrections significantly modify the tree-level effective potential that may prove useful for obtaining a needed amount of EDE. As already mentioned, this kind of attempt has been made in \cite{Maziashvili:2021mbm} for the model \eqref{example1}. When $\beta=0$, the cosmon field and CNB are uncoupled in all models cited above: Eq.\eqref{example1}, Eq.\eqref{example2}, Eq.\eqref{example3}, implying that in this case the one-loop contribution due to CNB to the effective potential \eqref{effpotential} should vanish. Demanding this normalization, the one-loop contribution due to CNB takes the form \cite{Kapusta:2006pm, Bailin:1986wt}

	\begin{eqnarray}\label{fermion}
	- 3 \int\frac{\mathrm{d}^3q}{(2\pi)^3} \left(\sqrt{\mathbf{q}^2+m_\nu^2(\phi)} -\sqrt{\mathbf{q}^2+\mu_\nu^2}\right) -      6T_\nu\int\frac{\mathrm{d}^3q}{(2\pi)^3} \ln \frac{1+\mathrm{e}^{-\left.\sqrt{\mathbf{q}^2+m_\nu^2(\phi)}\right/T_\nu}}{1+\mathrm{e}^{- \sqrt{\mathbf{q}^2+\mu_\nu^2}/T_\nu}}   ~. 
	\end{eqnarray} In the case of model \eqref{example1}, expanding the exponential functions gives an infinite number of even powers of $\phi$ and for removing divergences coming from quantum corrections \eqref{fermion} - the number of counter terms is also infinite \cite{Maziashvili:2021mbm}. That is, the model is only formally renormalizable. For this reason let us restrict our attention to the model \eqref{example3}. Separating the divergent part from the quantum corrections \eqref{fermion}

	\begin{align}
	&- 3 \int\frac{\mathrm{d}^3q}{(2\pi)^3} \left(\sqrt{\mathbf{q}^2+m_\nu^2(\phi)} -\sqrt{\mathbf{q}^2+\mu_\nu^2}\right)   = -3 \int \frac{\mathrm{d}^4 q}{(2\pi)^4}     \ln \frac{q^2+m_\nu^2(\phi)}{q^2+\mu_\nu^2} = \nonumber \\&   -\, \frac{3}{8\pi^2}\int_0^{~\, \Lambda} \mathrm{d}q \, q^3 \ln\left(\frac{q^2+m_\nu^2}{q^2+\mu_\nu^2}\right) = \nonumber \\&  -\, \frac{3}{32\pi^2} \left[\big(m_\nu^2 - \mu_\nu^2\big)\Lambda^2 -m_\nu^4\ln\frac{m_\nu^2+\Lambda^2}{m_\nu^2} + \mu_\nu^4\ln\frac{\mu_\nu^2+\Lambda^2}{\mu_\nu^2}  +\Lambda^4\ln \frac{m_\nu^2+\Lambda^2}{\Lambda^2} - \Lambda^4\ln \frac{\mu_\nu^2+\Lambda^2}{\Lambda^2} \right] = \nonumber \\&   -\, \frac{3}{32\pi^2}  \left[2\big(m_\nu^2 - \mu_\nu^2\big)\Lambda^2  - m_\nu^4 \left(\ln\frac{\Lambda^2}{m_\nu^2} +\frac{1}{2}\right)  + \mu_\nu^4\left(\ln\frac{\Lambda^2}{\mu_\nu^2} +\frac{1}{2}\right) + O\big(\Lambda^{-2}\big)  \right] = \nonumber \\& -\, \frac{3}{32\pi^2}  \left[2\big(m_\nu^2 - \mu_\nu^2\big)\Lambda^2 - \big(m_\nu^4 - \mu_\nu^4\big)\ln \frac{\Lambda^2}{\xi^2}  \right.  \left.  - m_\nu^4 \left(\ln\frac{\xi^2}{m_\nu^2} +\frac{1}{2}\right)  + \mu_\nu^4 \left(\ln\frac{\xi^2}{\mu_\nu^2} +\frac{1}{2}\right) + O\big(\Lambda^{-2}\big)  \right]  = \nonumber 
	\end{align}

	\begin{eqnarray}&&\label{correction1}  2\left(\frac{\beta^2\phi^2}{M_P^2} - \frac{2\beta \phi}{M_P}\right)\mu_\nu^2\Lambda^2 + \left( \frac{4\beta\phi}{M_P} - \frac{6\beta^2\phi^2}{M_P^2} +\frac{4\beta^3\phi^3}{M_P^3} \right.  \left. -\frac{\beta^4\phi^4}{M_P^4} \right)\mu_\nu^4\ln \frac{\Lambda^2}{\xi^2} - m_\nu^4 \left(\ln\frac{\xi^2}{m_\nu^2} +\frac{1}{2}\right)  + \nonumber \\&& \mu_\nu^4 \left(\ln\frac{\xi^2}{\mu_\nu^2} +\frac{1}{2}\right) + O\big(\Lambda^{-2}\big) ~, ~~~~~~~~~~
	\end{eqnarray} one can remove them by the finite number of counter terms for model \eqref{example3}. In view of the potential \eqref{example3}, the counter terms $\propto \phi, \phi^2, \phi^3$, that are dictated by Eq.\eqref{correction1} remove just the divergences and the corresponding renormalized quantities are put to zero. The subtraction point $\xi$ can be set for instance by the neutrino mass scale $\mu_\nu$. What we see from Eq.\eqref{correction1} is that the quantum corrections due to neutrinos are controlled by the neutrino mass scale ($m_\nu\leq$) $\mu_\nu $ and because of smallness of neutrino masses for obtaining needed amount of EDE that was discussed in the preceding section - this contribution to the EDE is obviously insufficient.

As for the thermal corrections \eqref{fermion}, the integral in convergent and in the high-temperature limit leads to th expression

\begin{eqnarray}
\frac{T^2_\nu(m^2_\nu-\mu_\nu^2)}{8} ~. \nonumber 
\end{eqnarray} Apart from the negative sign of this contribution, we see that its magnitude is obviously smaller than the required amount of EDE: $ 0.86T_\nu^4$.

Apart from the quantum corrections arising due to coupling with CNB, there is an additional contribution that
comes due to quantum fluctuations of $\phi$ field around the minimum, $\phi_+$, of the effective potential. This contribution is given by

\begin{eqnarray}\label{Nullpunktsenergie}
\frac{1}{2}\int\frac{\mathrm{d}^3 q}{(2\pi)^3} \left[\sqrt{\mathbf{q}^2 + m_{\text{eff}}^2(\phi)} -\sqrt{\mathbf{q}^2 + m_{\text{eff}}^2(0)} \right] ~, 
\end{eqnarray} where  \begin{eqnarray}
m_{\text{eff}}^2(\phi) \equiv  \mathfrak{U}''(\phi) =  3\lambda\phi^2 +\frac{\beta^2 \mu^2_\nu T^2_\nu}{4M_P^2}  ~. \label{effectivemass}
\end{eqnarray} We have chosen a specific normalization in \eqref{Nullpunktsenergie} ensuring that the quantum corrections are zero at $\phi=0$. The integral \eqref{Nullpunktsenergie} can be evaluated in much the same way as the first integral in Eq.\eqref{fermion}. Therefore, we arrive at the expression

\begin{eqnarray}\label{qcphi}
\frac{m_{\text{eff}}^4(\phi)}{64\pi^2}\ln \frac{m_{\text{eff}}^2(\phi)}{m_{\text{eff}}^2(0)} ~. 
\end{eqnarray} Now the question is to make the expression \eqref{qcphi} of the order of $0.86T_\nu^4$ in such a way as to keep the parameters $\lambda$ and $\beta \mu_\nu/M_P$ small for ensuring that the two-loop contributions are kept small. From Eq.\eqref{minimum} it follows that

\begin{eqnarray}
\lambda\phi_+^3 \,=\, \frac{\beta \mu^2_\nu T^2_\nu}{4M_P} \,-\, \frac{\beta^2\mu^2_\nu T^2_\nu}{4M_P^2}\phi_+ = \frac{\beta \mu^2_\nu T^2_\nu}{4M_P} \left(1 - \frac{1}{\eta}\right)  ~, \nonumber 
\end{eqnarray} where, for the sake of convenience we have introduced a parameter: $\beta\phi_+/M_P = 1/\eta$. For our purposes one should require $\eta\gg 1$ that implies that 

\begin{eqnarray}\label{validity}
\lambda\phi_+^3 \,\approx\, \frac{\beta \mu^2_\nu T^2_\nu}{4M_P} ~~\Rightarrow~~  \lambda\phi_+^2 \,=\, \eta \, \frac{\beta^2\mu^2_\nu T^2_\nu}{4M_P^2} ~, 
\end{eqnarray} and in order to make $m_{\text{eff}}(\phi_+)$ of the order of $T_\nu$ one should assume

\begin{eqnarray}
\eta \, \frac{\beta^2\mu^2_\nu T^2_\nu}{4M_P^2} \,\simeq\, T^2_\nu ~~\Rightarrow~~ \eta \,\simeq\, \frac{M_P^2}{\beta^2\mu^2_\nu} ~~\Rightarrow~~ \phi_+ \simeq \frac{\beta\mu^2_\nu}{M_P} ~. \nonumber 
\end{eqnarray} Unfortunately, the Eq.\eqref{validity} tells us that for obtaining this particular value of $\phi_+$ one should take 

\begin{eqnarray}
\lambda \,\simeq \, \frac{T^2_\nu}{\mu^2_\nu} \, \frac{M_P^2}{\beta^2\mu^2_\nu} \,\gg\, 1 . \nonumber 
\end{eqnarray}

Thus we arrive at the negative conclusion that it does not seem plausible to keep the (dimensionless) parameters $\lambda$ and $\beta \mu_\nu/M_P$ small (for disregarding two-loop corrections) and at the same time obtain a relatively large effective mass \eqref{effectivemass} that would ensure a needed amount of EDE in model \eqref{example3}.

\section{Additional remark.}

One may have an idea to save the model by admitting a solution that in addition to the potential energy $U(\phi_+)$ implies a non-negligible contribution $\dot{\phi}^2/2$ so that the total energy density provides the needed amount of EDE\footnote{The necessity of clarifying this point was pointed out to author by Referee.}. It is understood that this solution ensures the initial condition $\phi=0$ before the system approaches matter-radiation equality epoch. Upon approaching this epoch, the field begins to "thaw" and activate, exhibiting rapid oscillations around $\phi_+$ that subsequently dissipate by the friction term $H\dot{\phi}$. In favor of this idea, let us assume that the friction term can be neglected as the field undergoes rapid oscillations. The ignorance of this term makes the discussion simple and clear but of course it results in an overestimation of EDE. In order to extract robust results, we make one more simplifying assumption that the temperature $T_\nu$ varies slowly over the time-scale of these oscillations. The equation of motion

\begin{figure}[h]
	\centering
	\includegraphics[width=0.8\textwidth]{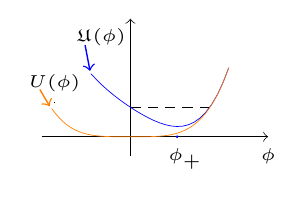}
	\caption{The schematic drawing depicting the potential and effective potential for the model under consideration. The dashed line marks the value of $\rho_\phi$. } \label{fig1}
\end{figure}

\begin{eqnarray}
	\ddot{\phi} =  -\partial_\phi \mathfrak{U}(\phi, T_\nu) ~, \nonumber  
\end{eqnarray} where $T_\nu$ is understood to be close to $1$\,eV, tells us that (see Fig.\ref{fig1})

\begin{eqnarray}\label{activation}
\rho_\phi  \equiv	\frac{\dot{\phi}^2}{2} +\frac{\lambda \phi^4}{4} = \frac{m_\nu^2(0)T_\nu^2}{8} =  \frac{\mu_\nu^2T_\nu^2}{8} ~. 
\end{eqnarray} From this estimate it is obvious that $\rho_\phi$ can not provide sufficient amount of EDE for $T_\nu \simeq 1$\,eV.  For obtaining the needed amount of EDE, one should take 

\begin{eqnarray}
	T_\nu \simeq \frac{\sqrt{8\times 0.86}}{0.05} \, \text{eV} \simeq 52 \, \text{eV} ~. \nonumber 
\end{eqnarray} This "thawing" temperature is by three orders of magnitude greater than the heaviest neutrino mass-eigenstate and maybe even higher if we recall that the above made evaluation overestimates the value of EDE. Again we see that because of smallness of neutrino masses one can not obtain the needed amount of EDE if the "activation temperature of the field" is somewhat close to the neutrino mass scale.

\section{Concluding remarks.}

For addressing two main challenges of dynamical dark energy (DE) models consisting of dynamical explanation of coincidence and the clarification of the question what makes DE to become predominant after the coincidence, one may find it preferable to consider coupled models of quintessence/cosmon field. One of the possibilities is to introduce the (conformal) coupling of cosmon to the cosmic neutrino background (CNB) \cite{Fardon:2003eh, Peccei:2004sz, Wetterich:2007kr, Amendola:2007yx, Brookfield:2005bz}. In the framework of this sort of models, one may hope to introduce the energy scale associated to DE density (at the time of coincidence) from particle physics sector and to dynamically provide the time-scale at which the coincidence or DE activation takes place. This time-scale is usually expected to be set by the time at which CNB enters a non-relativistic regime. Similar approach has been used in \cite{Sakstein:2019fmf, CarrilloGonzalez:2020oac} in attempt to relate the time-scale at which CNB gets nonrelativistic to the time of EDE activation. The latter occurs at the matter-radiation equality and is close to the time when $T_\nu$ becomes of the order of $1$\,eV. The role of EDE is to slightly increase the expansion rate of the universe prior to the recombination, while leaving the late universe unaltered \cite{Poulin:2018dzj, Agrawal:2019lmo, Smith:2019ihp}. In particular, the contribution of EDE to the total energy density of the universe should be about 10\% around the matter-radiation equality. Unfortunately, the mass-varying-neutrino model \cite{Sakstein:2019fmf, CarrilloGonzalez:2020oac} seems to be less efficient for EDE as it does not provide the needed amount of EDE density because of smallness of neutrino masses. Or to be more precise, looking at the result \eqref{activation}, one can argue that either the activation of $\phi$ occurs near $T_\nu\simeq 1$\,eV and the amount of EDE is insufficient or the activation of scalar field takes place well before the CNB approaches the non-relativistic regime and the activation of EDE cannot be related to the non-relativistic transition of CNB in any way. A supplementary discussion maybe found in \cite{deSouza:2023sqp}.

\appendix

\section{Appendix: Evaluating $\rho_\nu - 3p_\nu$ term.}
\label{appendixA}

In the case of free streaming neutrinos

\begin{eqnarray}
 \varepsilon_\nu(\mathbf{k}) = \sqrt{\frac{\mathbf{k}^2}{a^2}+m_\nu^2} ~, ~~ \rho_\nu = \frac{\mathsf{g}}{a^3}\int\frac{\mathrm{d}^3k}{(2\pi)^3}\, \frac{\varepsilon_\nu(\mathbf{k})}{\mathrm{e}^{k/aT_\nu} + 1}   ~,  ~~ p_\nu  =  \frac{\mathsf{g}}{3a^5}\int\frac{\mathrm{d}^3k}{(2\pi)^3}\frac{k^2}{\varepsilon_\nu(\mathbf{k})\Big(\mathrm{e}^{k/aT_\nu} + 1\Big)} ~, \nonumber 
\end{eqnarray} where the factor $\mathsf{g}$ accounts for the polarization degrees of freedom: two helicity states per flavor (or per mass-eigenstate). From these expressions one finds 

\begin{eqnarray}
\frac{\partial\rho_\nu}{\partial\phi} = 	\frac{\mathsf{g} m_\nu m_\nu'}{a^3} \int\frac{\mathrm{d}^3k}{(2\pi)^3}\, \frac{1}{\varepsilon(\mathbf{k})\left[\mathrm{e}^{k/aT_\nu} + 1\right]} ~,  ~~  \rho_\nu - 3p_\nu = 	\frac{\mathsf{g}m_\nu^2}{a^3} \int\frac{\mathrm{d}^3k}{(2\pi)^3}\, \frac{1}{\varepsilon_\nu(\mathbf{k})\left[\mathrm{e}^{k/aT_\nu} + 1\right]} ~,  \nonumber 
\end{eqnarray} and thus

\begin{eqnarray}\label{toloba}
\frac{m_\nu'}{m_\nu} \left[ \rho_\nu - 3p_\nu \right] = \frac{\partial\rho_\nu(\phi, T_\nu)}{\partial\phi} ~. 
\end{eqnarray} Let us note in passing that in view of equality \eqref{toloba}, the Eq.\eqref{continuity} reduces to

\begin{eqnarray}
\frac{\partial\rho_\nu(\phi, T_\nu)}{\partial T_\nu}\dot{T}_\nu + 3H \Big(\rho_\nu(\phi, T_\nu) + p_\nu(\phi, T_\nu)\Big) = 0 ~, \nonumber 
\end{eqnarray} that is automatically satisfied under assumption that $T_\nu\propto a^{-1}$. In general, it might be useful to replace \eqref{toloba} by

\begin{eqnarray}
\frac{m_\nu'}{m_\nu} \left[ \rho_\nu - 3p_\nu \right] = \frac{\partial\tilde{\rho}_\nu(\phi, T_\nu)}{\partial\phi} ~, \end{eqnarray} where   \begin{eqnarray}\tilde{\rho}_\nu(\phi, T_\nu) = \rho_\nu(\phi, T_\nu) - \rho_\nu(0, T_\nu) ~. 
\end{eqnarray} Changing the variable of integration from $k$ to $\xi = k/aT_\nu$, 

\begin{eqnarray}
\rho_\nu = \frac{\mathsf{g}}{2\pi^2 a^3}\int_0^\infty \mathrm{d}k \,  \frac{k^2\sqrt{k^2/a^2 +m_\nu^2}}{\mathrm{e}^{k/aT_\nu} + 1 } =  \frac{\mathsf{g} T_\nu^4}{2\pi^2 }\int_0^\infty \mathrm{d}\xi \, \frac{\xi^2\sqrt{\xi^2 +m_\nu^2/T_\nu^2}}{\mathrm{e}^\xi + 1 } ~, \nonumber 
\end{eqnarray} one finds approximate expressions

\begin{eqnarray}
\rho_\nu =  \frac{\mathsf{g} T_\nu^4}{2\pi^2 }  \left( \frac{7\pi^4}{120} + \frac{m_\nu^2\pi^2}{24 T^2}  \right) ~,  
\end{eqnarray} when $m_\nu/T_\nu \ll 1$ and

\begin{eqnarray}
\rho_\nu =  \frac{\mathsf{g}3\zeta(3) m_\nu T_\nu^3}{4\pi^2 } ~,
\end{eqnarray} when $m_\nu/T_\nu \gtrsim 1$.

\acknowledgments

The work was supported in part by the Shota Rustaveli National Science Foundation of Georgia under Grant No. FR-19-8306.


\end{document}